\documentclass[aps,pra,reprint,tightenlines,twocolumn,superscriptaddress,nofootinbib,showpacs]{revtex4-1}
\usepackage{amssymb}
\usepackage{color,graphicx}
\usepackage{lmodern}
\usepackage{amsmath}
\usepackage{amsthm}
\usepackage{bm}
\usepackage{float}
\usepackage{epstopdf}
\newcommand{\ket}[1]{\vert{#1}\rangle}
\newcommand{\bra}[1]{\langle{#1}\vert}

\newcommand{\proj}[1]{\ket{#1}\!\bra{#1}}

\newcommand{\one}{\openone}

\renewcommand\Re{\operatorname{Re }}
\renewcommand\Im{\operatorname{Im }}

\newcommand{\beq}{\begin{equation}}
\newcommand{\eeq}{\end{equation}}

\newcommand{\be}{\begin{equation}}
\newcommand{\ee}{\end{equation}}

\newcommand{\ben}{\begin{eqnarray}}
\newcommand{\een}{\end{eqnarray}}

\begin{document}
\title{Classicalization of Quantum State of Detector by Amplification Process}
\author{Arash Tirandaz}
\affiliation{Foundations of Physics Group, School of Physics, Institute for Research in Fundamental Sciences (IPM), P.O.Box 19395-5531, Tehran, Iran}
\author{Farhad Taher Ghahramani}
\email[Corresponding Author: ]{farhadtqm@ipm.ir}
\affiliation{Foundations of Physics Group, School of Physics, Institute for Research in Fundamental Sciences (IPM), P.O.Box 19395-5531, Tehran, Iran}
\author{Ali Asadian}
\affiliation{Department of Physics, Institute for Advanced Studies in Basic Sciences (IASBS), Gava Zang, Zanjan 45137-66731, Iran}
\affiliation{Naturwissenschaftlich-Technische Fakult\"at, Universit\"at Siegen, Walter-Flex-Str. 3, D-57068 Siegen, Germany}
\author{Mehdi Golshani}
\affiliation{Foundations of Physics Group, School of Physics, Institute for Research in Fundamental Sciences (IPM), P.O.Box 19395-5531, Tehran, Iran}
\affiliation{Department of Physics, Sharif University of Technology, P.O.Box 11365-9516, Tehran, Iran}

\begin{abstract}
It has been shown that a macroscopic system being in a high-temperature thermal coherent state can be, in principle, driven into a non-classical state by coupling to a microscopic system. Therefore, thermal coherent states do not truly represent the classical limit of quantum description. Here, we study the classical limit of quantum state of a more relevant macroscopic system, namely the pointer of a detector, after the phase-preserving linear amplification process. In particular, we examine to what extent it is possible to find the corresponding amplified state in a superposition state, by coupling the pointer to a qubit system. We demonstrate quantitatively that the amplification process is able to produce the classical limit of quantum state of the pointer, offering a route for a classical state in a sense of not to be projected into a quantum superposition state.
\end{abstract}
\maketitle
\section{Introduction}
Decoherence theory attempts to give an account of how classical behavior emerges from the quantum description by invoking the notion of environment~\cite{ZurekRMP,MaxRMP,Shah}. However, prior to that, the question is what is the classical limit of quantum state of a macroscopic  system\footnote{Note that we distinguish between `macroscopic' (`microscopic') and `classical' (`quantum'). The former refers to the system under study, the later to its description.} in the first place? As Bell said: ``What exactly qualifies some physical systems to play the role of measurer?"~\cite{Bell}. More relevant to the present work, we refine this question: what exactly qualifies a quantum state to represent the physical state of an isolated classical object, e.g. a measurement apparatus, or even a cat for that matter?\\
\indent Since the advent of quantum mechanics, there have been several proposals as to how the classical behavior emerges from the underling quantum dynamics (for a review see~\cite{Land}). The first one, proposed by Bohr as a {\it correspondence principle}, claimed that the classical behavior emerges in the limit of large quantum numbers~\cite{Bohr}. However, it turned out that this is not true in general. As a counterexample, a microscopic oscillator in an energy state with a large quantum number doesn't faithfully represent the classical behavior. This motivated Schr{\"o}dinger to propose an alternative quantum description of a ``classical-like" state of the oscillator, the so-called ``coherent state", whose dynamics closely resembles that of the classical one~\cite{Sch}.\\
\indent Despite the promising classical features of coherent states, because of the linearity of Schr{\"o}dinger equation, a system prepared in a coherent state can be driven into a superposition state by coupling it to a microscopic system~\cite{Schl,Knight,Chai,Janszky,Janszky2,Szabo,Vac1,Vac2,Joh1}. In fact, the non-classical features of coherent states are even recognized to be useful in quantum information science~\cite{Cir,Gros}. The problem remains even if we consider a more classical state; a high-temperature thermal coherent state~\cite{Joh2,Ralph,Ralph2,Zheng,Jeong}. It has been also demonstrated that such seemingly classical states violate the Leggett-Garg inequality~\cite{Asadian}. Therefore, it seems that the thermalization process by itself is not sufficient to classicalize a macroscopic system. \\
\indent In the early 1980s, it was suggested that the {\it amplification process} in detectors might have a crucial role in the quantum-to-classical transition~\cite{Mach1,Mach2,Ara,Nami}. The development of MASERs as possible amplifiers triggered a flurry of interest in the quantum description of amplification process in 1960s~\cite{Lou,Hef,Hau,Gor1,Gor2}, leading to the realization that a linear amplifier unavoidably adds noise to the input signal. This fundamental limit is expressed formally as a bound on the second moment of the added noise~\cite{Cav82,Clerk}. In this context, quantum non-demolition measurements are designed to circumvent the limitations imposed by such limit as repeated measurements of quantum states are performed~\cite{Bra,Cave80,Boc}. The quantum limit on the entire distribution of the added noise was provided just recently~\cite{Cave12}. Notably, a realistic amplifier transforms the quantum state of detector into a form which is not necessarily equivalent to the Gaussian distribution of the thermal coherent state. The amplified state is supposed to represent the pointer state from which the measurement result is read out, and thus we expect to have a definite value at the macroscopic scale. Therefore, we believe that it is timely to revisit this problem in the light of the recent advances in the amplifiers and see if the quantum description of a realistic amplifier yields the most ``classical-like" quantum state.\\
\indent In this work, we analyze the classicality of the pointer state after the amplification process, by looking at the possibility of projecting them into superposition states (see FIG.~\ref{QCircuit}). In particular, we focus on a mathematical model put forward by Caves and co-workers to examine phase-preserving linear amplifiers~\cite{Cave12}. In essence, the amplification of the input mode of pointer requires it to be coupled to an external mode, called ancillary mode, which adds noise to the output mode. The state of the ancillary mode determines the effect of the added noise on the input state. Our results show that a realistic i.e. non-ideal amplifier is able to produce classical states.
\section{Classical Limit of Quantum State}
\begin{figure}[H]
\centering
\includegraphics[width=0.4\textwidth]{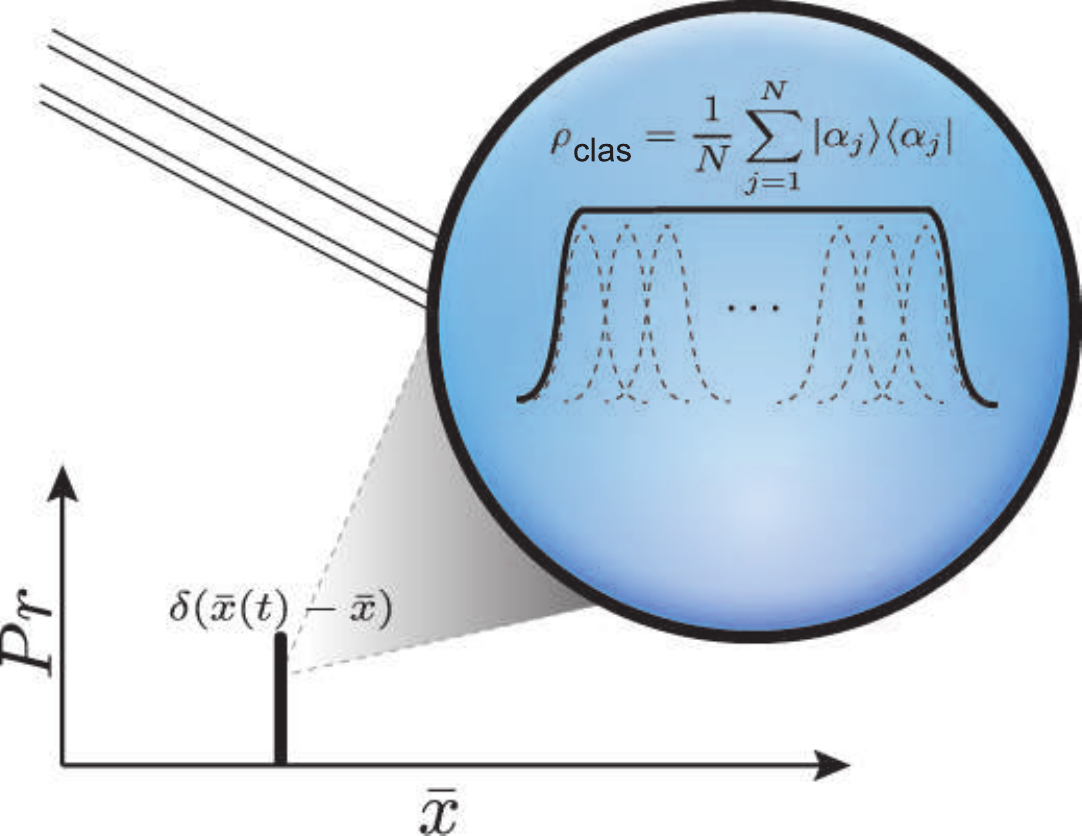}
\caption{The pointer's state, denoted by $\rho_{\rm clas}$, has a well-defined position $\bar x$ at the coarse-grained scale. However, a sharp measurement resolves different microscopic states which are consistent with the same pointer's macroscopic state.}
\label{Measurement}
\end{figure}

What is exactly the classical limit of quantum state of the pointer of a detector? Suppose that the pointer at the microscopic scale is prepared in a superposition of two energy states. According to the decoherence theory, given an appropriate interaction between the microscopic system and the surrounding environment, the pointer is found in an incoherent mixture of coherent states, e.g.
\begin{equation}\label{Coh}
\frac{1}{2}\proj{\alpha}+\frac{1}{2}\proj{-\alpha},
\end{equation}
which is widely considered as a classical mixture of distinct states. From quantum-mechanical point of view, however, it can be equally well represented by
\begin{equation}
\frac{1}{2}\proj{\psi^+}+\frac{1}{2}\proj{\psi^-}.
\end{equation}
where we defined $\ket{\psi^\pm}=\frac{1}{\sqrt{2}}(\ket{\alpha}\pm e^{i\varphi}\ket{-\alpha})$. A priorily quantum mechanics is unbiased toward any of these two inequivalent representations; as there is no unique ensemble decomposition of the mixed states. This so-called {\it basis ambiguity}~\cite{MaxRMP,Bassi} might look a mathematical problem, but it has physical consequences. Obviously, for $\bra{\alpha}-\alpha\rangle\approx 0$, a measurement in basis $\{\ket{\psi^\pm}\}$ results non-classical states\footnote{The same argument applies even to a thermal coherent state~\cite{Ralph}}. However, the representation of the pointer's state in all bases should be equivalent, because all dynamical operators of the pointer as a macroscopic system commute with each other. \\
\indent Form the above argument, we conclude that the classical state of the pointer is realized if two conditions are fulfilled. First,
if we expand the state in a given basis, there should be no interference between constituting states. This is a necessary but insufficient condition, because constituting states in another basis might have non-zero interference. This means that although there is no interference between constituting states, each can still be projected into a superposition state. Therefore, we add a second condition, asserting that there should be no interference for each constituting state too. \\
\indent To fulfill the second condition, the state should contain a sufficient amount of noise, hindering the generation of entanglement with a genuine microscopic quantum state, and yet maintaining a well-defined pointer position in the coarse-grained macroscopic scale, as expected from a macroscopic pointer. In fact, the classical state does not reveal the detailed quantum state of the pointer. This means that there might be many quantum states, in the fine-grained scale, basically producing the same classical state of the pointer. Therefore, the classical state of the pointer is represented by a probabilistic mixture of all these consistent quantum states. Here, since the classical state does not distinguish between the micro-states, we assign equal probability to them. This produces a top-flat probability distribution (see FIG.~\ref{Measurement}). We define the corresponding classical state as
\begin{equation}\label{N}
\rho_{\rm clas}=\frac{1}{N}\sum_{j=1}^{N}\proj{\alpha_{j}}.
\end{equation}
In the appendix, we demonstrated how such a noisy, top-flat distribution destroys the interference in all bases. \\
Such an ansatz has been justified in different contexts. Kofler and Brukner suggested that ``coarse-grained measurements'' give rise to classical behavior~\cite{Kofler}. Also, Zhuang and co-workers employ a similar notion, called continuous variable ``information scrambling'' to characterize macroscopic systems~\cite{Yao}. Scrambling refers to the dynamical delocalization of quantum information over an entire system's degrees of freedom. More relevant to our work, Rossi and co-workers demonstrated that the classical limit of a macroscopic system is mathematically well-defined, if the system exhibits some global symmetry~\cite{Rossi}. In the short-time regime, it is the symmetry under the permutation of different modes, which is realized as ``narrow energy spectrum'' condition. This  condition is reflected as a ``top-flat'' distribution in our work. \\
\indent But how the classical state (\ref{N}) is realized in the first place? The recent advances on quantum description of realistic amplifiers, motivated us to look at the classicality of amplified states. Amplification process not only amplifies the input state at a price of adding noise, producing a macroscopic state which is noisy at the fine-grained microscopic scale, and yet having well-defined pointer position at the coarse-grained macroscopic scale. It is worth mentioning that the resulting amplified state is not  equivalent to a thermal coherent state. Therefore, it is intriguing to see if, unlike a thermal coherent state, its probability of being found in a superposition state, vanishes. Before investigating this idea, let us briefly review the amplification process.
\section{Amplification Process}
\begin{figure}[H]
\includegraphics[width=0.8\columnwidth]{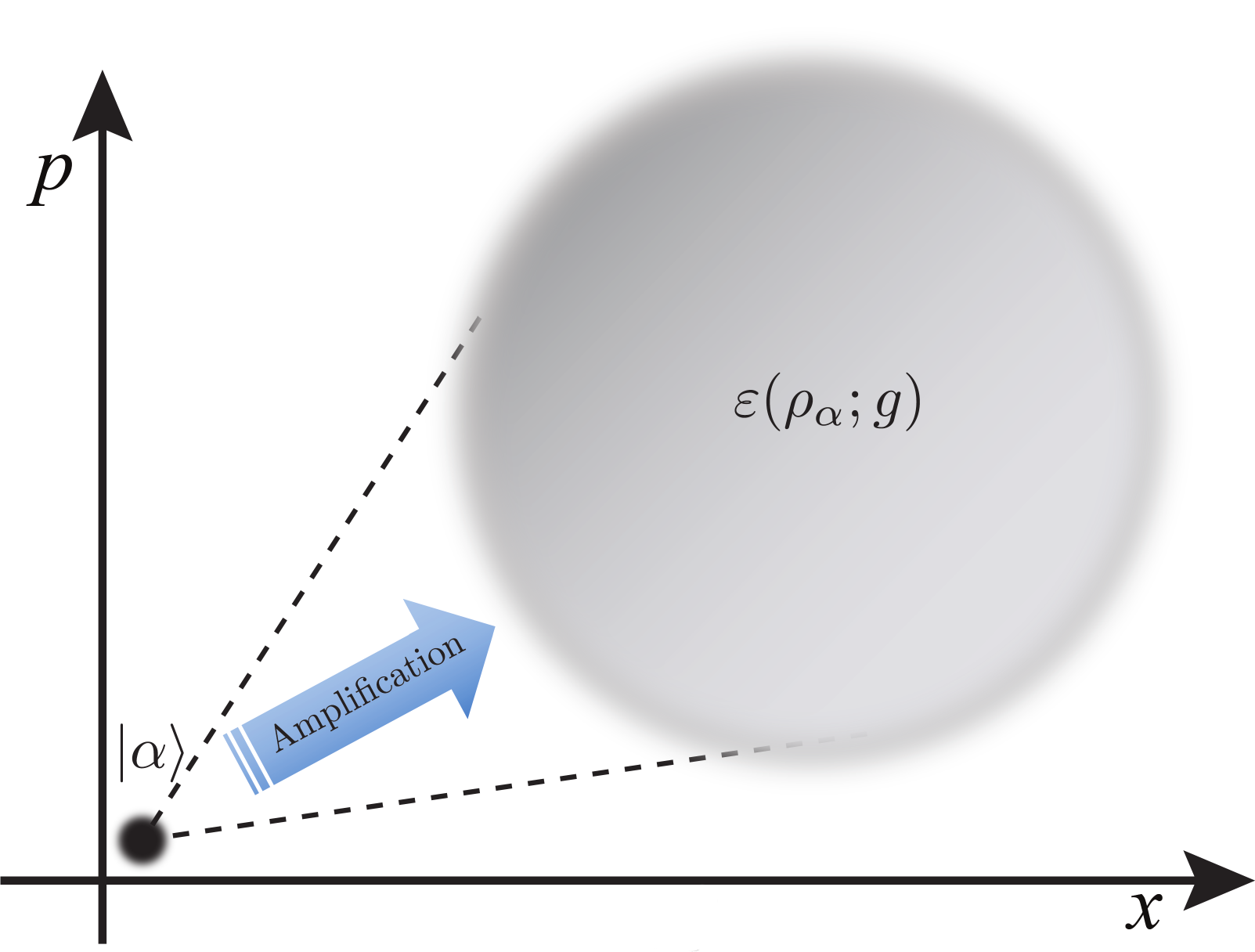}
\caption{The initial coherent state undergoes linear phase-preserving amplification, resulting the smearing of the probability distribution in phase space.}
\label{Amplification}
\end{figure}

The setting for our analysis is a bosonic mode $\hat a$, called {\it primary} mode, which is to undergo amplification process. The type of amplification one typically thinks of in physics is {\it linear} amplification, which means that the output mode is linearly related to the input mode (here, being multiplied by some fixed amplitude gain $g$). Here, we shall deal with linear amplification, since it is straightforward to treat mathematically. Also we wish to amplify both quadratures of the input mode with the same gain. This type of amplification is often referred to as {\it phase-preserving} amplification (FIG.~\ref{Amplification}).
\noindent A {\it perfect} phase-preserving linear amplifier transforms the input mode directly to the output one: $\hat a_{out}=g\hat a_{in}$. However, this transformation violates unitarity. Physically, this means that amplification of the primary mode requires it to be coupled to an external mode $\hat b$, called ancillary mode, which adds noise to the output mode. When referred to the input, the output noise is constrained as $\langle|\Delta a_{out}|^{2}\rangle/g^{2}\geq\langle|\Delta a_{in}|^{2}\rangle+1/2$~\cite{Cav82}. The minimum added noise, corresponding to the lower bound, is the half-quantum of vacuum noise. The amplifier working with the minimum added noise is called an {\it ideal} amplifier. The simplest model of such an amplifier is provided by a parametric amplifier~\cite{Mol1,Mol2,Col}. The ideal amplified state of the input state $\rho$ is given by~\cite{Cave12}
\begin{equation}
\varepsilon(\rho;g)=tr_{b}(\hat S\rho\otimes\sigma\hat S^{\dagger}),
\end{equation}
where $\hat S=e^{r(\hat a\hat b-\hat a^{\dagger}\hat b^{\dagger})}$ is the two-mode squeezing operator, with the amplitude gain being $g=\cosh r$, and $\sigma$ is the positive density operator of the ancillary mode. The density operator $\sigma$ is diagonal in the number basis
\begin{equation}
\sigma=\sum_{n=0}^{\infty}\lambda_{n}|n\rangle\langle n|,
\end{equation}
where `$\lambda_{n}$'s are the corresponding eigenvalues. Note that for an ideal amplifier, we have $\sigma=|0\rangle\langle 0|$. \\
\indent The amplified state for the complete distribution of the added noise is given by the amplifier map~\cite{Cave12}
\begin{equation}
\varepsilon(\rho;g)=\hat B\big(\hat A(g)\rho\big).
\end{equation}
The superoperator $\hat A$ amplifies the input state $\rho$ with the gain $g$. For a coherent input state, the output of $\hat A(g)$ is just a displaced coherent state: $\hat A(g)|\alpha\rangle\langle\alpha|=|g\alpha\rangle\langle g\alpha|$. The superoperator $\hat B$ adds a noise to the output state by smearing out a phase-space distribution into a broader distribution as
\begin{equation}
\hat B=\int d^{2}\beta~\Pi^{-1}(\beta)\hat D(\hat a,\beta)\odot\hat D^{\dagger}(\hat a,\beta),
\end{equation}
where $\odot$ marks the slot where the input to the superoperator goes and $\hat D(\hat a,\beta)$ is the displacement operator for the mode $\hat a$; $\hat D(\hat a,\beta)=e^{\beta\hat a^{\dagger}-\beta^{\ast}\hat a}$. The real-valued function $\Pi^{-1}(\beta)$ is called the {\it smearing function}. This function is independent of the input state, but it depends on the gain $g$ as
\begin{equation}
\Pi^{-1}(\alpha)=\frac{e^{-|\alpha|^{2}/(g^{2}-1)}}{\pi(g^{2}-1)}\sum_{n=0}^{\infty}\frac{\lambda_{n}|\alpha|^{2n}}{n!(g^{2}-1)^{n}}.
\end{equation}
\noindent Note that the smearing function of an ideal linear amplifier is isomorphic to the Glauber-Sudarshan P function of the thermal coherent state $P_{th}(v,d)=\frac{2}{\pi(v-1)}\exp[-\frac{2|\alpha-d|^{2}}{v-1}]$, where $v$ is the variance and $d$ is the displacement in the phase-space. We safely assume that before amplification due to the internal dynamics of the macroscopic system, the system is evolved to the most classical pure state, i.e. a coherent state $\rho_{\alpha}=|\alpha\rangle\langle\alpha|$.\\
\indent Our analysis here is mathematically based on the optical states. Nonetheless, the  generalization of our approach to the corresponding mechanical states is straightforward. For example, a superconducting qubit can manifest macroscopic distinguishable states by injecting currents of opposite verses. The corresponding wave functions are effectively Gaussian states. They are isomorphic to the coherent states in our analysis.
\section{A case study: The pointer interacting with a two-level system}
\begin{figure}[H]
\includegraphics[width=1\columnwidth]{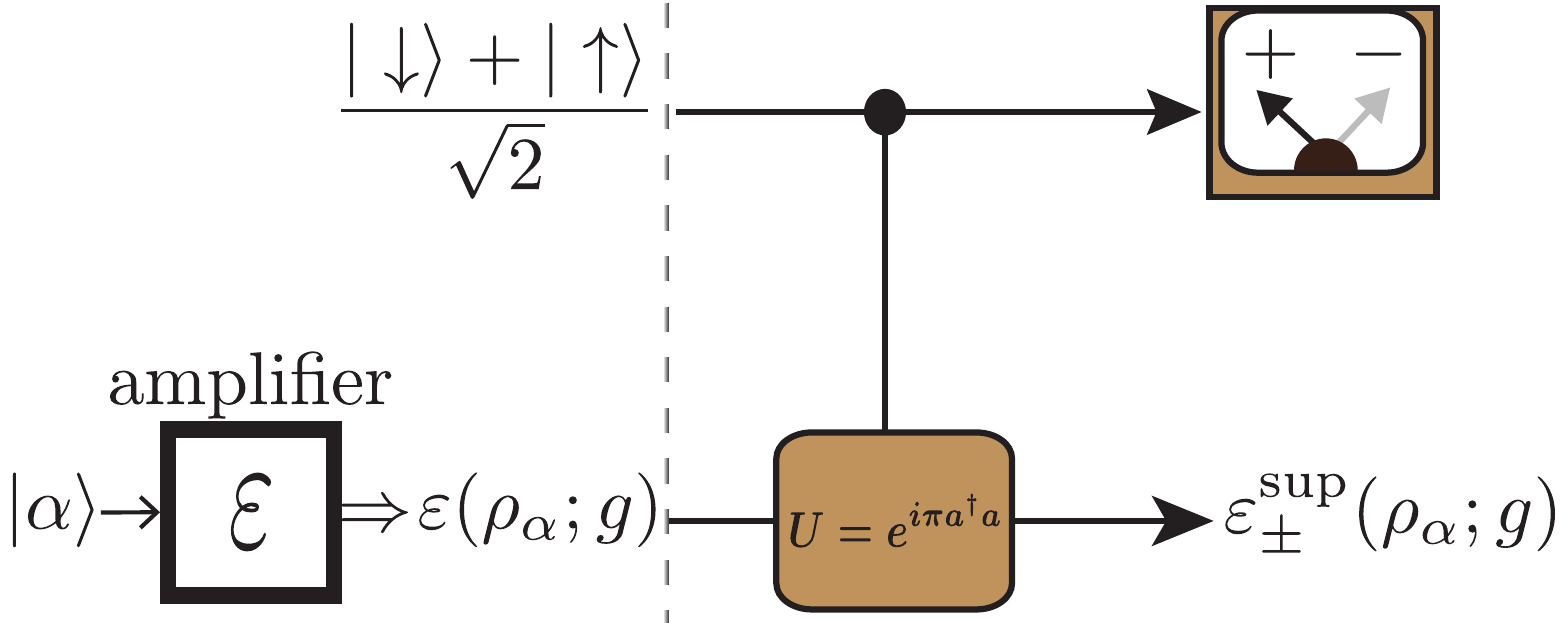}
\caption{Circuit for projecting the amplified state to a non-classical state, by coupling it to a two-level system.}
\label{QCircuit}
\end{figure}

\begin{figure}[H]
\centering
\includegraphics[scale=0.1]{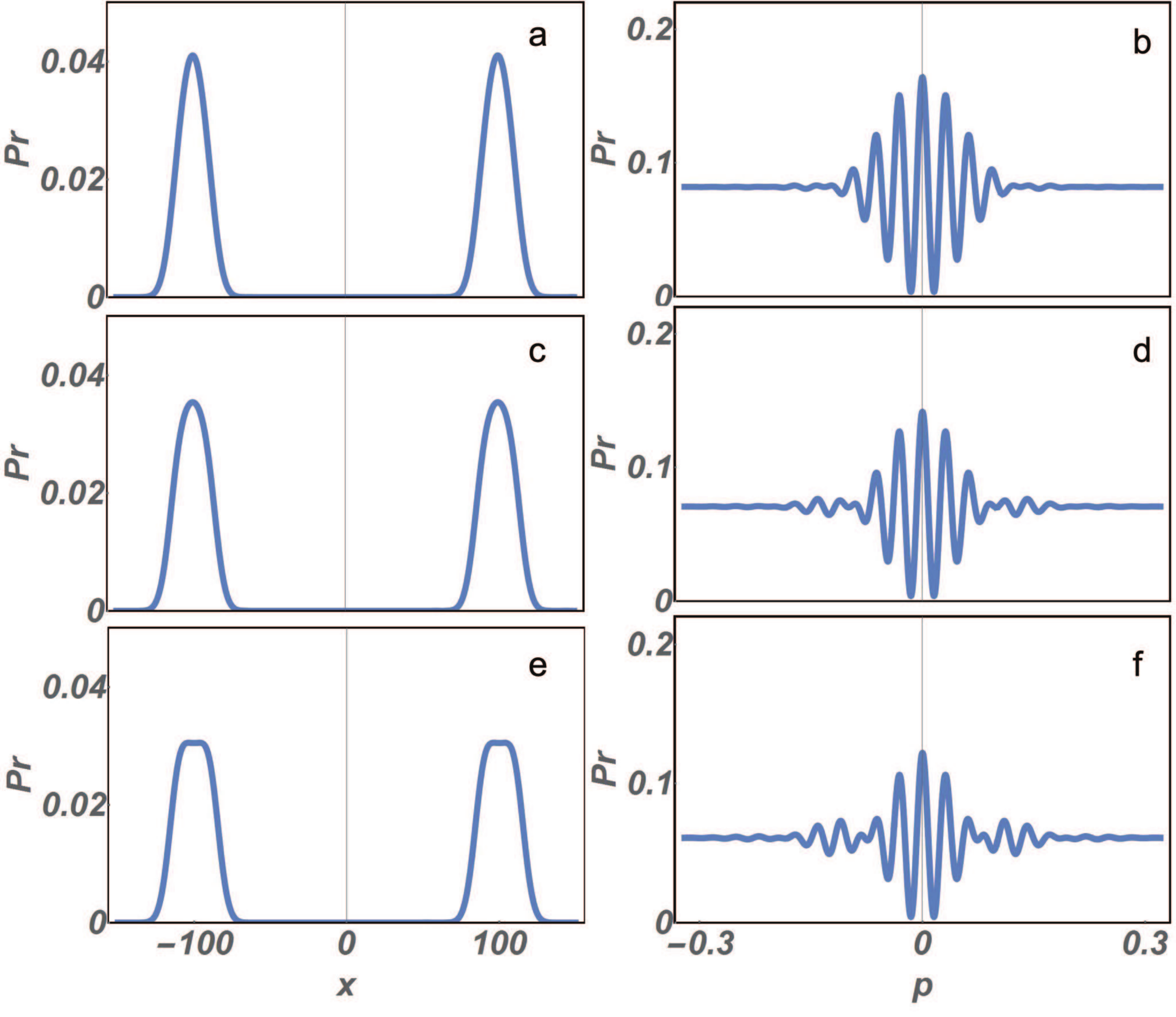}
\caption{The probability of $x$ (left) and probability of $p$ (right) for an amplified coherent state with $g=\alpha=10$ for the first three terms with $\lambda_{1}=0.5$ , $\lambda_{2}=0.3$ and $\lambda_{3}=0.2$ (a,b), $\lambda_{1}=0.33$ , $\lambda_{2}=0.33$ and $\lambda_{3}=0.33$ (c,d), and $\lambda_{1}=0.2$ , $\lambda_{2}=0.3$ and $\lambda_{3}=0.5$ (e,f). The noisy, top-flat distribution with the minimum of interference is obtained by an increasing order of eigenvalues $\lambda_{n}$.}
\label{Landas}
\end{figure}

\begin{figure}[h]
\centering
\includegraphics[scale=0.85]{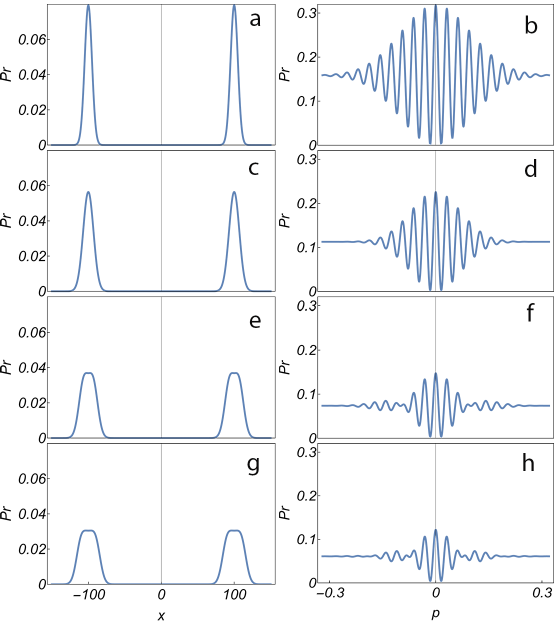}
\caption{The probability of $x$ (left) and probability of $p$ (right) for a high-temperature thermal coherent state with $d=v=100$ (a,b), an amplified coherent state with $g=\alpha=10$ for the first term (c,d), for the first two terms with $\lambda_{1}=0.3$, $\lambda_{2}=0.7$ (e,f), and for the first three terms with $\lambda_{1}=0.2$, $\lambda_{2}=0.3$ and $\lambda_{3}=0.5$ (g,h). It is obvious that as we add the non-ideal effects to the ideal linear amplifier, the interference fringes are weakened gradually.}
\label{Basic}
\end{figure}

\begin{figure}[h]
\centering
\includegraphics[width=0.45\textwidth]{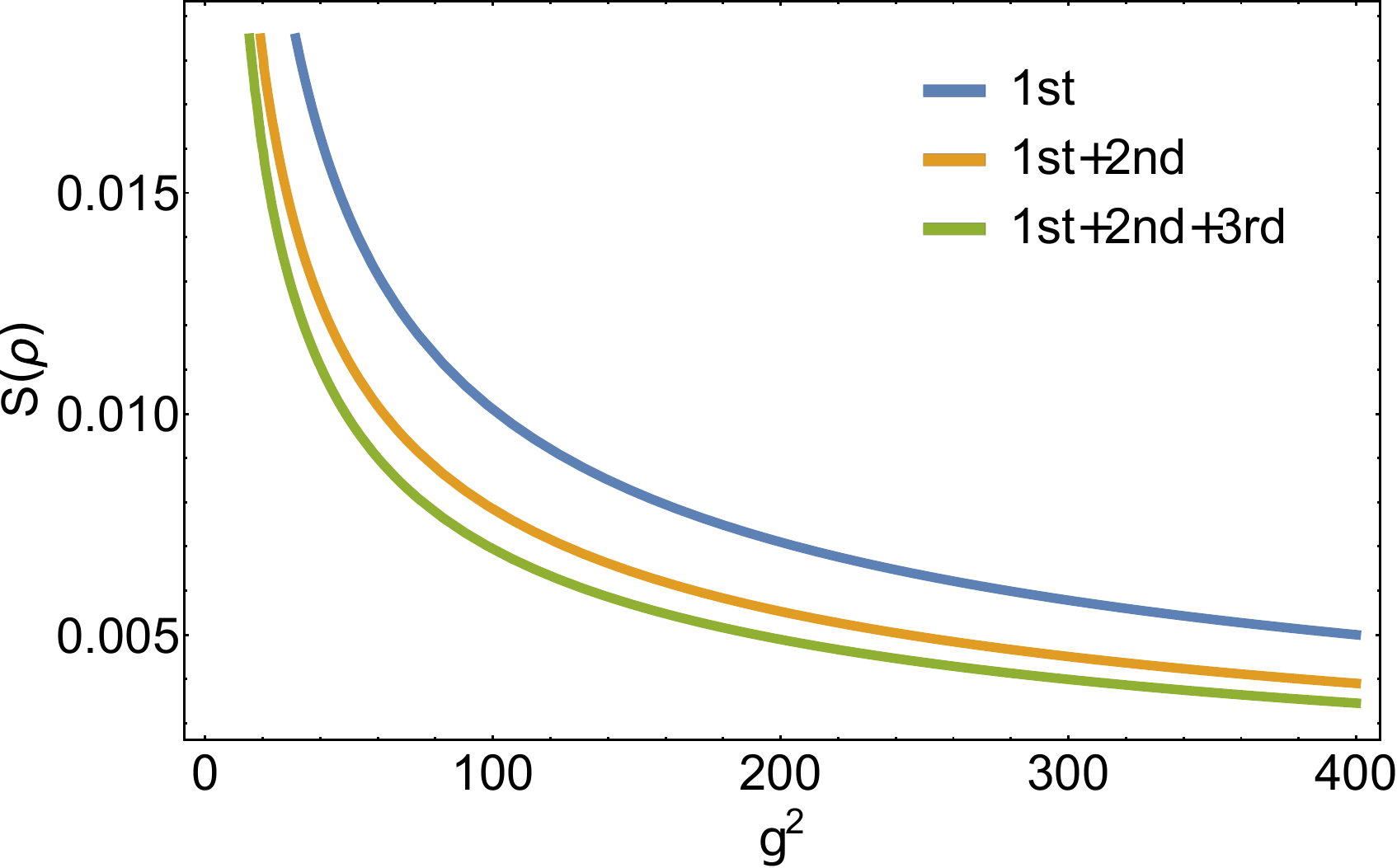}
\caption{Interference-base measure $S(\rho)$ for a linear amplifier with $\alpha=10$ for the first term (blue), for two first terms with $\lambda_{1}=0.3$, $\lambda_{2}=0.7$ (orange), for three first terms with $\lambda_{1}=0.2$, $\lambda_{2}=0.3$, $\lambda_{3}=0.5$ (green). This shows that the non-ideal linear amplifier with large gain $g$ is able to produce a classical state.}
\label{Meas} 
\end{figure}

\begin{figure}[h]
\centering
\includegraphics[scale=0.64]{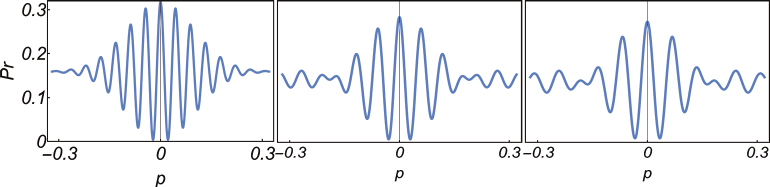}
\caption{The probability of $p$ for an amplified coherent state including the first term with $g=7.10$ (left), the first two terms with $\lambda_{1}=0.3$, $\lambda_{2}=0.7$ and $g=5.28$ (middle) and the first three terms with $\lambda_{1}=0.2$, $\lambda_{2}=0.3$, $\lambda_{3}=0.5$ and $g=4.56$ (right). The purity of these amplified states is equal to the purity of a thermal coherent state, with $v=d=100$, which is $0.01$.}
\label{Purity}
\end{figure}

Now we check whether an amplified state is classical in the sense we defined in section II. The state of the pointer after amplification is given by
\begin{equation}
\varepsilon(\rho_{\alpha};g)=\int d^{2}\beta~\Pi^{-1}(\beta-g\alpha)|\beta\rangle\langle\beta|.
\end{equation}
We consider $\varepsilon(\rho_{\alpha};g)$ as the initial state of the pointer, interacting with a qubit system, as illustrated in Fig.~\ref{QCircuit}.
\noindent The qubit is prepared in state \mbox{$|+\rangle=(|\uparrow\rangle+|\downarrow\rangle)/\sqrt{2}$}, where  $|\uparrow\rangle$ and $|\downarrow\rangle$ are the spin-up and spin-down states of the qubit in the $z$ direction. Qubit being in $\ket{+}$ interacts with the pointer prepared in the amplified coherent state $\varepsilon(\rho_{\alpha};g)$, with the interaction Hamiltonian being $c\hat a^{\dagger}\hat a|\uparrow\rangle\langle\uparrow|$, where $c$ is the coupling strength. After $t=\pi/c$, the resulting state is given by
\begin{align}
\label{Ent}
&U_\pi\Big(\proj{+}\otimes\varepsilon(\rho_{\alpha};g)\Big)U_\pi^\dag\\ \nonumber
&=\proj{+}\otimes E_+\varepsilon(\rho_{\alpha};g)E_+^\dag+\proj{-}\otimes E_-\varepsilon(\rho_{\alpha};g)E_-^\dag \\
\nonumber
&+\ket{+}\bra{-}\otimes E_+\varepsilon(\rho_{\alpha};g)E_-^\dag+\ket{-}\bra{+}\otimes E_-\varepsilon(\rho_{\alpha};g)E_+^\dag.
\end{align}
where $E_\pm=(\one\pm U_\pi)/2$ and $U_\pi=\exp(i\pi a^\dag a)$. Upon the qubit measurement on the basis $|\pm\rangle$, the detector's state is projected into a superposition state
\begin{align}
\varepsilon^{\rm sup}_{\pm}(\rho_{\alpha};g)=\frac{E_\pm\varepsilon(\rho_{\alpha};g)E_\pm^\dag}{p_\pm}
\end{align}
with
\begin{align}
E_\pm\varepsilon&(\rho_{\alpha};g)E_\pm^\dag=\!\int\! d^{2}\beta~\Pi^{-1}(\beta-g\alpha)\nonumber\\&
\big\{|\beta\rangle\langle\beta|+|-\beta\rangle\langle-\beta|\pm|\beta\rangle\langle-\beta|\pm|-\beta\rangle\langle\beta|\big\},
\end{align}
and
\begin{equation}
p_\pm={\rm Tr}\{E_\pm\varepsilon(\rho_{\alpha};g)E_\pm^\dag\}
\end{equation}
is the probability of finding the amplifier in the corresponding superposition state, $\varepsilon^{\rm sup}_{\pm}(\rho_{\alpha};g)$.\\
\indent The probability distribution of diagonal and off-diagonal elements of $\varepsilon^{sup}(\rho_{\alpha};g)$ can be obtained as $Pr(x)=\langle x|\varepsilon^{sup}(\rho_{\alpha};g)|x\rangle$ and $Pr(p)=\langle p|\varepsilon^{sup}(\rho_{\alpha};g)|p\rangle$, respectively. The two peaks along $x(\equiv\Re\alpha)$ axis are well-separated and represent the pointer positions, if the measurement has been performed in $\{\ket{\uparrow}, \ket{\downarrow}\}$ basis. Interference fringes along $p(\equiv\Im\alpha)$ axis are a typical signature of quantum superposition between macroscopically distinct states. It is worth mentioning that the interference pattern indicates the generated quantum entanglement between the qubit and the amplified mode proceeding the projective measurement on the qubit. Therefore, any amplified state truly representing the classical limit should suppress the generated entanglement with the qubit. The entanglement being suppressed, the pointer is now classically correlated to the system being measured, i.e. the state of the total system in (\ref{Ent}) reduces to its first two terms. \\
\indent The amplitude and the pattern of peaks and interference fringes depend on the choices of ancillary eigenvalues $\lambda_{n}$. The only constraint imposed by quantum mechanics is to guarantee that $\sigma$ is a valid density operator, $\lambda_{n}$s should be non-negative\footnote{Carlton M. Caves, Private Communication.}. It is not yet clear in detail how $\lambda_n$s are parameterized in actual ``non-ideal" amplifiers. Nonetheless, we found that the most appropriate choice to ensure the emergence of classicality is $0<\lambda_{n}<\lambda_{n+1}$ (see Fig.~\ref{Landas}). Note that the visibility can be always unity. 
\indent The peaks and interference fringes for a high-temperature thermal coherent state and the corresponding amplified coherent states for ancillary mode with one, two and three available states are plotted in FIG.~\ref{Basic}, for an optimized choice of $\lambda_{n}$s.
\noindent Notably, the non-ideal amplification process for certain range of parameters produces a probability distribution (see FIG.~\ref{Basic}(e,g)) which has similar top-flat shape with that of our heuristic model illustrated in FIG.~\ref{Measurement}. \\
\indent According to FIG.~\ref{Basic}, as we include the non-ideal terms in the ideal phase-preserving linear amplifier, the interference effects vanish gradually. This also can be verified, using a quantitative measure of macroscopicity. Lee and Jeong introduced a general and inclusive measure of macroscopicity in the phase space as~\cite{Lee}
\begin{equation}
S(\rho)\!=\!\frac{\pi^{M}}{2}\!\int\! d^{2}\alpha W(\alpha)\sum_{m=1}^{M}\big[-\frac{\partial^{2}}{\partial\alpha_{m}\partial\alpha^{\ast}_{m}}-1\big]W(\alpha),
\end{equation}
where $W(\alpha)$ is the Wigner function of the state and $M$ is the number of modes. The measure $S(\rho)$ was plotted for amplified coherent states with the first term, with  first two terms and with  first three terms in FIG.~\ref{Meas}.
As we expected, this shows that with a large gain $g$, as we include the non-ideal terms in the ideal amplifier, the probability of macroscopic quantum superpositions decreases.\\
\indent To ensure that the non-ideal linear amplifier has a non-thermal effect, we compare the interference fringes appeared in the high-temperature thermal coherent state with those of a non-ideal amplified coherent state with equal purity, ${\rm Tr}(\rho_{th}^2)={\rm Tr}(\{\varepsilon(\rho_\alpha;g)\}^2)$. For a thermal coherent state with $v=d=100$, the amplifier gain $g$ of the corresponding amplified coherent state with the first term, the first two terms and the first three terms are $7.10$, $5.28$ and $4.56$, respectively. For a fixed purity, the interference fringes for an amplifier with the corresponding terms were plotted in FIG.~\ref{Purity}. The suppression of interference shows that including non-ideal terms has a non-thermal effect. The inclusion of many of terms in the actual detector can vanish the interference.
\section{Discussion \& concluding remarks}
It is of significance to come up with a form of quantum state representing the state of a macroscopic system, and resolving the problem of basis ambiguity, without necessarily invoking the entanglement with an environment or modifying the laws of quantum mechanics. That is why, we look at the amplifier, as it is supposed to describe the pointer's state, and examine the basis ambiguity of the amplified state via a non-classical evolution sketched in FIG.~\ref{QCircuit}. Our results show that an amplifier state can demonstrate a behavior which is not similar to that of a thermal coherent state when it is coupled to a microscopic state, yet having well-defined pointer position at the coarse-grained macroscopic scale. Nonetheless, we stress that to achieve a more conclusive result we need to examine a non-ideal amplifier state with realistic parameters. To our best knowledge, this is yet to be fully identified, and then this issue requires further research.
\section*{Acknowledgment}
A. Asadian acknowledges Consolidator Grant 683107/TempoQ.
\appendix*
\section{Resolution of Basis Ambiguity for Classical State}
\setcounter{equation}{0}
Here we prove that the classical state (\ref{N}) is the same in all bases. In order to make more connection to the classical phase space description, we shift to the Wigner representation
\begin{align}
W(\alpha)&=\frac{1}{\pi^{2}N}\sum_{j}\int d^{2}\beta\langle\alpha_{j}|D(\beta)|\alpha_{j}\rangle e^{-(\beta\alpha^{\ast}-\beta^{\ast}\alpha)}\nonumber \\&=\frac{2}{\pi N}\sum_{j}e^{-2|\alpha-\alpha_{j}|^{2}},
\end{align}
where $D(\beta)$ is the displacement operator. As the summation is over all coherent states which lies in a single slot $\bar\alpha$ in the coarse-grained scale, we have
\begin{equation}
W(\alpha)=\delta(\alpha-\bar\alpha),
\end{equation}
where $\bar\alpha$ is a coarse-grained phase amplitude which has classical dynamics. The mixed state corresponding to (\ref{Coh}) would then be
\begin{equation}\label{Old}
\frac{1}{2}\delta(\alpha-\bar\alpha)+\frac{1}{2}\delta(\alpha+\bar\alpha).
\end{equation}
We may construct a new basis in terms of the old one as
\begin{align}
\rho^{\pm}=\frac{4}{N}\sum_{j}^{N}&\Big(|\alpha_{j}\rangle\langle\alpha_{j}|+|-\alpha_{j}\rangle\langle-\alpha_{j}|
\nonumber \\&\pm|-\alpha_{j}\rangle\langle\alpha_{j}|\pm|\alpha_{j}\rangle\langle-\alpha_{j}|\Big).
\end{align}
The mixed state (\ref{Old}) can be equivalently decomposed in the new basis as
\begin{equation}\label{TChi}
\frac{1}{2}\rho^{+}+\frac{1}{2}\rho^{-},
\end{equation}
where the corresponding Wigner representation is
\begin{align}\label{New}
\frac{1}{2}\delta(\alpha-\bar\alpha)&+\frac{1}{2}\delta(\alpha+\bar\alpha)\nonumber\\&\pm\frac{2^{-2|\alpha|^{2}}}{\pi^{2}N}\sum_{i}\cos(4Im(\alpha_{i}^{\ast}\alpha)).
\end{align}
The interference term is negligible, thus the decomposition in the new basis (\ref{New}) is approximately the decomposition in the old basis (\ref{Old}). This means that the classicalized state of the pointer leads to a unique ensemble decomposition at coarse-grained scale. Therefore, $\delta(\alpha-\bar\alpha)$ and $\delta(\alpha+\bar\alpha)$ emerge dynamically as the unique coarse-grained or classical preferred basis.\\


\begin{thebibliography}{47}
\makeatletter
\providecommand \@ifxundefined [1]{%
 \@ifx{#1\undefined}
}%
\providecommand \@ifnum [1]{%
 \ifnum #1\expandafter \@firstoftwo
 \else \expandafter \@secondoftwo
 \fi
}%
\providecommand \@ifx [1]{%
 \ifx #1\expandafter \@firstoftwo
 \else \expandafter \@secondoftwo
 \fi
}%
\providecommand \natexlab [1]{#1}%
\providecommand \enquote  [1]{``#1''}%
\providecommand \bibnamefont  [1]{#1}%
\providecommand \bibfnamefont [1]{#1}%
\providecommand \citenamefont [1]{#1}%
\providecommand \href@noop [0]{\@secondoftwo}%
\providecommand \href [0]{\begingroup \@sanitize@url \@href}%
\providecommand \@href[1]{\@@startlink{#1}\@@href}%
\providecommand \@@href[1]{\endgroup#1\@@endlink}%
\providecommand \@sanitize@url [0]{\catcode `\\12\catcode `\$12\catcode
  `\&12\catcode `\#12\catcode `\^12\catcode `\_12\catcode `\%12\relax}%
\providecommand \@@startlink[1]{}%
\providecommand \@@endlink[0]{}%
\providecommand \url  [0]{\begingroup\@sanitize@url \@url }%
\providecommand \@url [1]{\endgroup\@href {#1}{\urlprefix }}%
\providecommand \urlprefix  [0]{URL }%
\providecommand \Eprint [0]{\href }%
\providecommand \doibase [0]{http://dx.doi.org/}%
\providecommand \selectlanguage [0]{\@gobble}%
\providecommand \bibinfo  [0]{\@secondoftwo}%
\providecommand \bibfield  [0]{\@secondoftwo}%
\providecommand \translation [1]{[#1]}%
\providecommand \BibitemOpen [0]{}%
\providecommand \bibitemStop [0]{}%
\providecommand \bibitemNoStop [0]{.\EOS\space}%
\providecommand \EOS [0]{\spacefactor3000\relax}%
\providecommand \BibitemShut  [1]{\csname bibitem#1\endcsname}%
\let\auto@bib@innerbib\@empty
\bibitem [{\citenamefont {Zurek}(2003)}]{ZurekRMP}%
  \BibitemOpen
  \bibfield  {author} {\bibinfo {author} {\bibfnamefont {W.~H.}\ \bibnamefont
  {Zurek}},\ }\href {\doibase 10.1103/RevModPhys.75.715} {\bibfield  {journal}
  {\bibinfo  {journal} {Rev. Mod. Phys.}\ }\textbf {\bibinfo {volume} {75}},\
  \bibinfo {pages} {715} (\bibinfo {year} {2003})}\BibitemShut {NoStop}%
\bibitem [{\citenamefont {Schlosshauer}(2005)}]{MaxRMP}%
  \BibitemOpen
  \bibfield  {author} {\bibinfo {author} {\bibfnamefont {M.}~\bibnamefont
  {Schlosshauer}},\ }\href {\doibase 10.1103/RevModPhys.76.1267} {\bibfield
  {journal} {\bibinfo  {journal} {Rev. Mod. Phys.}\ }\textbf {\bibinfo {volume}
  {76}},\ \bibinfo {pages} {1267} (\bibinfo {year} {2005})}\BibitemShut
  {NoStop}%
\bibitem [{\citenamefont {Shahandeh}\ \emph {et~al.}(2017)\citenamefont
  {Shahandeh}, \citenamefont {Costa}, \citenamefont {Lund},\ and\ \citenamefont
  {Ralph}}]{Shah}%
  \BibitemOpen
  \bibfield  {author} {\bibinfo {author} {\bibfnamefont {F.}~\bibnamefont
  {Shahandeh}}, \bibinfo {author} {\bibfnamefont {F.}~\bibnamefont {Costa}},
  \bibinfo {author} {\bibfnamefont {A.~P.}\ \bibnamefont {Lund}}, \ and\
  \bibinfo {author} {\bibfnamefont {T.~C.}\ \bibnamefont {Ralph}},\ }\href@noop
  {} {\bibfield  {journal} {\bibinfo  {journal} {arXiv preprint
  arXiv:1711.10498}\ } (\bibinfo {year} {2017})}\BibitemShut {NoStop}%
\bibitem [{\citenamefont {Bell}(2004)}]{Bell}%
  \BibitemOpen
  \bibfield  {author} {\bibinfo {author} {\bibfnamefont {J.~S.}\ \bibnamefont
  {Bell}},\ }\href@noop {} {\emph {\bibinfo {title} {Speakable and unspeakable
  in quantum mechanics}}}\ (\bibinfo  {publisher} {Cambridge University
  Press},\ \bibinfo {year} {2004})\BibitemShut {NoStop}%
\bibitem [{\citenamefont {Landsman}(2006)}]{Land}%
  \BibitemOpen
  \bibfield  {author} {\bibinfo {author} {\bibfnamefont {N.~P.}\ \bibnamefont
  {Landsman}},\ }\href@noop {} {\bibfield  {journal} {\bibinfo  {journal}
  {Handbook of the Philosophy of Science}\ }\textbf {\bibinfo {volume} {2}},\
  \bibinfo {pages} {417} (\bibinfo {year} {2006})}\BibitemShut {NoStop}%
\bibitem [{\citenamefont {Bohr}(1923)}]{Bohr}%
  \BibitemOpen
  \bibfield  {author} {\bibinfo {author} {\bibfnamefont {N.}~\bibnamefont
  {Bohr}},\ }\href@noop {} {\bibfield  {journal} {\bibinfo  {journal}
  {Zeitschrift f{\"u}r Physik}\ }\textbf {\bibinfo {volume} {13}},\ \bibinfo
  {pages} {117} (\bibinfo {year} {1923})}\BibitemShut {NoStop}%
\bibitem [{\citenamefont {Schr{\"o}dinger}(1926)}]{Sch}%
  \BibitemOpen
  \bibfield  {author} {\bibinfo {author} {\bibfnamefont {E.}~\bibnamefont
  {Schr{\"o}dinger}},\ }\href@noop {} {\bibfield  {journal} {\bibinfo
  {journal} {Naturwissenschaften}\ }\textbf {\bibinfo {volume} {14}},\ \bibinfo
  {pages} {664} (\bibinfo {year} {1926})}\BibitemShut {NoStop}%
\bibitem [{\citenamefont {Schleich}\ \emph {et~al.}(1991)\citenamefont
  {Schleich}, \citenamefont {Pernigo},\ and\ \citenamefont {Le~Kien}}]{Schl}%
  \BibitemOpen
  \bibfield  {author} {\bibinfo {author} {\bibfnamefont {W.}~\bibnamefont
  {Schleich}}, \bibinfo {author} {\bibfnamefont {M.}~\bibnamefont {Pernigo}}, \
  and\ \bibinfo {author} {\bibfnamefont {F.}~\bibnamefont {Le~Kien}},\
  }\href@noop {} {\bibfield  {journal} {\bibinfo  {journal} {Phys. Rev. A}\
  }\textbf {\bibinfo {volume} {44}},\ \bibinfo {pages} {2172} (\bibinfo {year}
  {1991})}\BibitemShut {NoStop}%
\bibitem [{\citenamefont {Bu\ifmmode~\check{z}\else \v{z}\fi{}ek}\ \emph
  {et~al.}(1992)\citenamefont {Bu\ifmmode~\check{z}\else \v{z}\fi{}ek},
  \citenamefont {Vidiella-Barranco},\ and\ \citenamefont {Knight}}]{Knight}%
  \BibitemOpen
  \bibfield  {author} {\bibinfo {author} {\bibfnamefont {V.}~\bibnamefont
  {Bu\ifmmode~\check{z}\else \v{z}\fi{}ek}}, \bibinfo {author} {\bibfnamefont
  {A.}~\bibnamefont {Vidiella-Barranco}}, \ and\ \bibinfo {author}
  {\bibfnamefont {P.~L.}\ \bibnamefont {Knight}},\ }\href {\doibase
  10.1103/PhysRevA.45.6570} {\bibfield  {journal} {\bibinfo  {journal} {Phys.
  Rev. A}\ }\textbf {\bibinfo {volume} {45}},\ \bibinfo {pages} {6570}
  (\bibinfo {year} {1992})}\BibitemShut {NoStop}%
\bibitem [{\citenamefont {Chai}(1992)}]{Chai}%
  \BibitemOpen
  \bibfield  {author} {\bibinfo {author} {\bibfnamefont {C.-l.}\ \bibnamefont
  {Chai}},\ }\href {\doibase 10.1103/PhysRevA.46.7187} {\bibfield  {journal}
  {\bibinfo  {journal} {Phys. Rev. A}\ }\textbf {\bibinfo {volume} {46}},\
  \bibinfo {pages} {7187} (\bibinfo {year} {1992})}\BibitemShut {NoStop}%
\bibitem [{\citenamefont {Janszky}\ \emph {et~al.}(1993)\citenamefont
  {Janszky}, \citenamefont {Domokos},\ and\ \citenamefont {Adam}}]{Janszky}%
  \BibitemOpen
  \bibfield  {author} {\bibinfo {author} {\bibfnamefont {J.}~\bibnamefont
  {Janszky}}, \bibinfo {author} {\bibfnamefont {P.}~\bibnamefont {Domokos}}, \
  and\ \bibinfo {author} {\bibfnamefont {P.}~\bibnamefont {Adam}},\ }\href
  {\doibase 10.1103/PhysRevA.48.2213} {\bibfield  {journal} {\bibinfo
  {journal} {Phys. Rev. A}\ }\textbf {\bibinfo {volume} {48}},\ \bibinfo
  {pages} {2213} (\bibinfo {year} {1993})}\BibitemShut {NoStop}%
\bibitem [{\citenamefont {Janszky}\ \emph {et~al.}(1995)\citenamefont
  {Janszky}, \citenamefont {Domokos}, \citenamefont {Szab\'o},\ and\
  \citenamefont {Adam}}]{Janszky2}%
  \BibitemOpen
  \bibfield  {author} {\bibinfo {author} {\bibfnamefont {J.}~\bibnamefont
  {Janszky}}, \bibinfo {author} {\bibfnamefont {P.}~\bibnamefont {Domokos}},
  \bibinfo {author} {\bibfnamefont {S.}~\bibnamefont {Szab\'o}}, \ and\
  \bibinfo {author} {\bibfnamefont {P.}~\bibnamefont {Adam}},\ }\href {\doibase
  10.1103/PhysRevA.51.4191} {\bibfield  {journal} {\bibinfo  {journal} {Phys.
  Rev. A}\ }\textbf {\bibinfo {volume} {51}},\ \bibinfo {pages} {4191}
  (\bibinfo {year} {1995})}\BibitemShut {NoStop}%
\bibitem [{\citenamefont {Szabo}\ \emph {et~al.}(1996)\citenamefont {Szabo},
  \citenamefont {Adam}, \citenamefont {Janszky},\ and\ \citenamefont
  {Domokos}}]{Szabo}%
  \BibitemOpen
  \bibfield  {author} {\bibinfo {author} {\bibfnamefont {S.}~\bibnamefont
  {Szabo}}, \bibinfo {author} {\bibfnamefont {P.}~\bibnamefont {Adam}},
  \bibinfo {author} {\bibfnamefont {J.}~\bibnamefont {Janszky}}, \ and\
  \bibinfo {author} {\bibfnamefont {P.}~\bibnamefont {Domokos}},\ }\href
  {\doibase 10.1103/PhysRevA.53.2698} {\bibfield  {journal} {\bibinfo
  {journal} {Phys. Rev. A}\ }\textbf {\bibinfo {volume} {53}},\ \bibinfo
  {pages} {2698} (\bibinfo {year} {1996})}\BibitemShut {NoStop}%
\bibitem [{\citenamefont {Wiseman}\ and\ \citenamefont
  {Vaccaro}(2002{\natexlab{a}})}]{Vac1}%
  \BibitemOpen
  \bibfield  {author} {\bibinfo {author} {\bibfnamefont {H.~M.}\ \bibnamefont
  {Wiseman}}\ and\ \bibinfo {author} {\bibfnamefont {J.~A.}\ \bibnamefont
  {Vaccaro}},\ }\href@noop {} {\bibfield  {journal} {\bibinfo  {journal} {Phys.
  Rev. A}\ }\textbf {\bibinfo {volume} {65}},\ \bibinfo {pages} {043605}
  (\bibinfo {year} {2002}{\natexlab{a}})}\BibitemShut {NoStop}%
\bibitem [{\citenamefont {Wiseman}\ and\ \citenamefont
  {Vaccaro}(2002{\natexlab{b}})}]{Vac2}%
  \BibitemOpen
  \bibfield  {author} {\bibinfo {author} {\bibfnamefont {H.~M.}\ \bibnamefont
  {Wiseman}}\ and\ \bibinfo {author} {\bibfnamefont {J.~A.}\ \bibnamefont
  {Vaccaro}},\ }\href@noop {} {\bibfield  {journal} {\bibinfo  {journal} {Phy.
  Rev. A}\ }\textbf {\bibinfo {volume} {65}},\ \bibinfo {pages} {043606}
  (\bibinfo {year} {2002}{\natexlab{b}})}\BibitemShut {NoStop}%
\bibitem [{\citenamefont {Johansen}(2004{\natexlab{a}})}]{Joh1}%
  \BibitemOpen
  \bibfield  {author} {\bibinfo {author} {\bibfnamefont {L.~M.}\ \bibnamefont
  {Johansen}},\ }\href@noop {} {\bibfield  {journal} {\bibinfo  {journal}
  {Phys. Let. A}\ }\textbf {\bibinfo {volume} {329}},\ \bibinfo {pages} {184}
  (\bibinfo {year} {2004}{\natexlab{a}})}\BibitemShut {NoStop}%
\bibitem [{\citenamefont {Cirel'son}(1980)}]{Cir}%
  \BibitemOpen
  \bibfield  {author} {\bibinfo {author} {\bibfnamefont {B.~S.}\ \bibnamefont
  {Cirel'son}},\ }\href@noop {} {\bibfield  {journal} {\bibinfo  {journal}
  {Lett. Math. Phys.}\ }\textbf {\bibinfo {volume} {4}},\ \bibinfo {pages} {93}
  (\bibinfo {year} {1980})}\BibitemShut {NoStop}%
\bibitem [{\citenamefont {Grosshans}\ \emph {et~al.}(2003)\citenamefont
  {Grosshans}, \citenamefont {Van~Assche}, \citenamefont {Wenger},
  \citenamefont {Brouri}, \citenamefont {Cerf},\ and\ \citenamefont
  {Grangier}}]{Gros}%
  \BibitemOpen
  \bibfield  {author} {\bibinfo {author} {\bibfnamefont {F.}~\bibnamefont
  {Grosshans}}, \bibinfo {author} {\bibfnamefont {G.}~\bibnamefont
  {Van~Assche}}, \bibinfo {author} {\bibfnamefont {J.}~\bibnamefont {Wenger}},
  \bibinfo {author} {\bibfnamefont {R.}~\bibnamefont {Brouri}}, \bibinfo
  {author} {\bibfnamefont {N.~J.}\ \bibnamefont {Cerf}}, \ and\ \bibinfo
  {author} {\bibfnamefont {P.}~\bibnamefont {Grangier}},\ }\href@noop {}
  {\bibfield  {journal} {\bibinfo  {journal} {Nature}\ }\textbf {\bibinfo
  {volume} {421}},\ \bibinfo {pages} {238} (\bibinfo {year}
  {2003})}\BibitemShut {NoStop}%
\bibitem [{\citenamefont {Johansen}(2004{\natexlab{b}})}]{Joh2}%
  \BibitemOpen
  \bibfield  {author} {\bibinfo {author} {\bibfnamefont {L.~M.}\ \bibnamefont
  {Johansen}},\ }\href@noop {} {\bibfield  {journal} {\bibinfo  {journal} {J.
  Opt. B: Quantum Semiclassical Opt.}\ }\textbf {\bibinfo {volume} {6}},\
  \bibinfo {pages} {L21} (\bibinfo {year} {2004}{\natexlab{b}})}\BibitemShut
  {NoStop}%
\bibitem [{\citenamefont {Jeong}\ and\ \citenamefont {Ralph}(2006)}]{Ralph}%
  \BibitemOpen
  \bibfield  {author} {\bibinfo {author} {\bibfnamefont {H.}~\bibnamefont
  {Jeong}}\ and\ \bibinfo {author} {\bibfnamefont {T.~C.}\ \bibnamefont
  {Ralph}},\ }\href {\doibase 10.1103/PhysRevLett.97.100401} {\bibfield
  {journal} {\bibinfo  {journal} {Phys. Rev. Lett.}\ }\textbf {\bibinfo
  {volume} {97}},\ \bibinfo {pages} {100401} (\bibinfo {year}
  {2006})}\BibitemShut {NoStop}%
\bibitem [{\citenamefont {Jeong}\ and\ \citenamefont {Ralph}(2007)}]{Ralph2}%
  \BibitemOpen
  \bibfield  {author} {\bibinfo {author} {\bibfnamefont {H.}~\bibnamefont
  {Jeong}}\ and\ \bibinfo {author} {\bibfnamefont {T.~C.}\ \bibnamefont
  {Ralph}},\ }\href {\doibase 10.1103/PhysRevA.76.042103} {\bibfield  {journal}
  {\bibinfo  {journal} {Phys. Rev. A}\ }\textbf {\bibinfo {volume} {76}},\
  \bibinfo {pages} {042103} (\bibinfo {year} {2007})}\BibitemShut {NoStop}%
\bibitem [{\citenamefont {Zheng}(2007)}]{Zheng}%
  \BibitemOpen
  \bibfield  {author} {\bibinfo {author} {\bibfnamefont {S.-B.}\ \bibnamefont
  {Zheng}},\ }\href {\doibase 10.1103/PhysRevA.75.032114} {\bibfield  {journal}
  {\bibinfo  {journal} {Phys. Rev. A}\ }\textbf {\bibinfo {volume} {75}},\
  \bibinfo {pages} {032114} (\bibinfo {year} {2007})}\BibitemShut {NoStop}%
\bibitem [{\citenamefont {Jeong}\ \emph {et~al.}(2008)\citenamefont {Jeong},
  \citenamefont {Lee},\ and\ \citenamefont {Nha}}]{Jeong}%
  \BibitemOpen
  \bibfield  {author} {\bibinfo {author} {\bibfnamefont {H.}~\bibnamefont
  {Jeong}}, \bibinfo {author} {\bibfnamefont {J.}~\bibnamefont {Lee}}, \ and\
  \bibinfo {author} {\bibfnamefont {H.}~\bibnamefont {Nha}},\ }\href {\doibase
  10.1364/JOSAB.25.001025} {\bibfield  {journal} {\bibinfo  {journal} {J. Opt.
  Soc. Am. B}\ }\textbf {\bibinfo {volume} {25}},\ \bibinfo {pages} {1025}
  (\bibinfo {year} {2008})}\BibitemShut {NoStop}%
\bibitem [{\citenamefont {Asadian}\ \emph {et~al.}(2014)\citenamefont
  {Asadian}, \citenamefont {Brukner},\ and\ \citenamefont {Rabl}}]{Asadian}%
  \BibitemOpen
  \bibfield  {author} {\bibinfo {author} {\bibfnamefont {A.}~\bibnamefont
  {Asadian}}, \bibinfo {author} {\bibfnamefont {C.}~\bibnamefont {Brukner}}, \
  and\ \bibinfo {author} {\bibfnamefont {P.}~\bibnamefont {Rabl}},\ }\href
  {\doibase 10.1103/PhysRevLett.112.190402} {\bibfield  {journal} {\bibinfo
  {journal} {Phys. Rev. Lett.}\ }\textbf {\bibinfo {volume} {112}},\ \bibinfo
  {pages} {190402} (\bibinfo {year} {2014})}\BibitemShut {NoStop}%
\bibitem [{\citenamefont {Machida}\ and\ \citenamefont
  {Namiki}(1980{\natexlab{a}})}]{Mach1}%
  \BibitemOpen
  \bibfield  {author} {\bibinfo {author} {\bibfnamefont {S.}~\bibnamefont
  {Machida}}\ and\ \bibinfo {author} {\bibfnamefont {M.}~\bibnamefont
  {Namiki}},\ }\href@noop {} {\bibfield  {journal} {\bibinfo  {journal} {Prog.
  Theor. Phys.}\ }\textbf {\bibinfo {volume} {63}},\ \bibinfo {pages} {1457}
  (\bibinfo {year} {1980}{\natexlab{a}})}\BibitemShut {NoStop}%
\bibitem [{\citenamefont {Machida}\ and\ \citenamefont
  {Namiki}(1980{\natexlab{b}})}]{Mach2}%
  \BibitemOpen
  \bibfield  {author} {\bibinfo {author} {\bibfnamefont {S.}~\bibnamefont
  {Machida}}\ and\ \bibinfo {author} {\bibfnamefont {M.}~\bibnamefont
  {Namiki}},\ }\href@noop {} {\bibfield  {journal} {\bibinfo  {journal} {Prog.
  Theor. Phys.}\ }\textbf {\bibinfo {volume} {63}},\ \bibinfo {pages} {1833}
  (\bibinfo {year} {1980}{\natexlab{b}})}\BibitemShut {NoStop}%
\bibitem [{\citenamefont {Araki}(1980)}]{Ara}%
  \BibitemOpen
  \bibfield  {author} {\bibinfo {author} {\bibfnamefont {H.}~\bibnamefont
  {Araki}},\ }\href@noop {} {\bibfield  {journal} {\bibinfo  {journal} {Progr.
  Theor. Phys.}\ }\textbf {\bibinfo {volume} {64}},\ \bibinfo {pages} {719}
  (\bibinfo {year} {1980})}\BibitemShut {NoStop}%
\bibitem [{\citenamefont {Namiki}\ \emph {et~al.}(1998)\citenamefont {Namiki},
  \citenamefont {Nakazato},\ and\ \citenamefont {Pascazio}}]{Nami}%
  \BibitemOpen
  \bibfield  {author} {\bibinfo {author} {\bibfnamefont {M.}~\bibnamefont
  {Namiki}}, \bibinfo {author} {\bibfnamefont {H.}~\bibnamefont {Nakazato}}, \
  and\ \bibinfo {author} {\bibfnamefont {S.}~\bibnamefont {Pascazio}},\
  }\href@noop {} {\emph {\bibinfo {title} {Decoherence and quantum
  measurements}}}\ (\bibinfo  {publisher} {World Scientific},\ \bibinfo {year}
  {1998})\BibitemShut {NoStop}%
\bibitem [{\citenamefont {Louisell}\ \emph {et~al.}(1961)\citenamefont
  {Louisell}, \citenamefont {Yariv},\ and\ \citenamefont {Siegman}}]{Lou}%
  \BibitemOpen
  \bibfield  {author} {\bibinfo {author} {\bibfnamefont {W.}~\bibnamefont
  {Louisell}}, \bibinfo {author} {\bibfnamefont {A.}~\bibnamefont {Yariv}}, \
  and\ \bibinfo {author} {\bibfnamefont {A.}~\bibnamefont {Siegman}},\
  }\href@noop {} {\bibfield  {journal} {\bibinfo  {journal} {Phys. Rev.}\
  }\textbf {\bibinfo {volume} {124}},\ \bibinfo {pages} {1646} (\bibinfo {year}
  {1961})}\BibitemShut {NoStop}%
\bibitem [{\citenamefont {Heffner}(1962)}]{Hef}%
  \BibitemOpen
  \bibfield  {author} {\bibinfo {author} {\bibfnamefont {H.}~\bibnamefont
  {Heffner}},\ }\href@noop {} {\bibfield  {journal} {\bibinfo  {journal}
  {Proceedings of the IRE}\ }\textbf {\bibinfo {volume} {50}},\ \bibinfo
  {pages} {1604} (\bibinfo {year} {1962})}\BibitemShut {NoStop}%
\bibitem [{\citenamefont {Haus}\ and\ \citenamefont {Mullen}(1962)}]{Hau}%
  \BibitemOpen
  \bibfield  {author} {\bibinfo {author} {\bibfnamefont {H.~A.}\ \bibnamefont
  {Haus}}\ and\ \bibinfo {author} {\bibfnamefont {J.}~\bibnamefont {Mullen}},\
  }\href@noop {} {\bibfield  {journal} {\bibinfo  {journal} {Phys. Rev.}\
  }\textbf {\bibinfo {volume} {128}},\ \bibinfo {pages} {2407} (\bibinfo {year}
  {1962})}\BibitemShut {NoStop}%
\bibitem [{\citenamefont {Gordon}\ \emph
  {et~al.}(1963{\natexlab{a}})\citenamefont {Gordon}, \citenamefont {Walker},\
  and\ \citenamefont {Louisell}}]{Gor1}%
  \BibitemOpen
  \bibfield  {author} {\bibinfo {author} {\bibfnamefont {J.}~\bibnamefont
  {Gordon}}, \bibinfo {author} {\bibfnamefont {L.}~\bibnamefont {Walker}}, \
  and\ \bibinfo {author} {\bibfnamefont {W.}~\bibnamefont {Louisell}},\
  }\href@noop {} {\bibfield  {journal} {\bibinfo  {journal} {Phys. Rev.}\
  }\textbf {\bibinfo {volume} {130}},\ \bibinfo {pages} {806} (\bibinfo {year}
  {1963}{\natexlab{a}})}\BibitemShut {NoStop}%
\bibitem [{\citenamefont {Gordon}\ \emph
  {et~al.}(1963{\natexlab{b}})\citenamefont {Gordon}, \citenamefont
  {Louisell},\ and\ \citenamefont {Walker}}]{Gor2}%
  \BibitemOpen
  \bibfield  {author} {\bibinfo {author} {\bibfnamefont {J.}~\bibnamefont
  {Gordon}}, \bibinfo {author} {\bibfnamefont {W.~H.}\ \bibnamefont
  {Louisell}}, \ and\ \bibinfo {author} {\bibfnamefont {L.}~\bibnamefont
  {Walker}},\ }\href@noop {} {\bibfield  {journal} {\bibinfo  {journal} {Phys.
  Rev.}\ }\textbf {\bibinfo {volume} {129}},\ \bibinfo {pages} {481} (\bibinfo
  {year} {1963}{\natexlab{b}})}\BibitemShut {NoStop}%
\bibitem [{\citenamefont {Caves}(1982)}]{Cav82}%
  \BibitemOpen
  \bibfield  {author} {\bibinfo {author} {\bibfnamefont {C.~M.}\ \bibnamefont
  {Caves}},\ }\href {\doibase 10.1103/PhysRevD.26.1817} {\bibfield  {journal}
  {\bibinfo  {journal} {Phys. Rev. D}\ }\textbf {\bibinfo {volume} {26}},\
  \bibinfo {pages} {1817} (\bibinfo {year} {1982})}\BibitemShut {NoStop}%
\bibitem [{\citenamefont {Clerk}\ \emph {et~al.}(2010)\citenamefont {Clerk},
  \citenamefont {Devoret}, \citenamefont {Girvin}, \citenamefont {Marquardt},\
  and\ \citenamefont {Schoelkopf}}]{Clerk}%
  \BibitemOpen
  \bibfield  {author} {\bibinfo {author} {\bibfnamefont {A.~A.}\ \bibnamefont
  {Clerk}}, \bibinfo {author} {\bibfnamefont {M.~H.}\ \bibnamefont {Devoret}},
  \bibinfo {author} {\bibfnamefont {S.~M.}\ \bibnamefont {Girvin}}, \bibinfo
  {author} {\bibfnamefont {F.}~\bibnamefont {Marquardt}}, \ and\ \bibinfo
  {author} {\bibfnamefont {R.~J.}\ \bibnamefont {Schoelkopf}},\ }\href
  {\doibase 10.1103/RevModPhys.82.1155} {\bibfield  {journal} {\bibinfo
  {journal} {Rev. Mod. Phys.}\ }\textbf {\bibinfo {volume} {82}},\ \bibinfo
  {pages} {1155} (\bibinfo {year} {2010})}\BibitemShut {NoStop}%
\bibitem [{\citenamefont {Braginskii}(1988)}]{Bra}%
  \BibitemOpen
  \bibfield  {author} {\bibinfo {author} {\bibfnamefont {V.~B.}\ \bibnamefont
  {Braginskii}},\ }\href@noop {} {\bibfield  {journal} {\bibinfo  {journal}
  {Physics-Uspekhi}\ }\textbf {\bibinfo {volume} {31}},\ \bibinfo {pages} {836}
  (\bibinfo {year} {1988})}\BibitemShut {NoStop}%
\bibitem [{\citenamefont {Caves}\ \emph {et~al.}(1980)\citenamefont {Caves},
  \citenamefont {Thorne}, \citenamefont {Drever}, \citenamefont {Sandberg},\
  and\ \citenamefont {Zimmermann}}]{Cave80}%
  \BibitemOpen
  \bibfield  {author} {\bibinfo {author} {\bibfnamefont {C.~M.}\ \bibnamefont
  {Caves}}, \bibinfo {author} {\bibfnamefont {K.~S.}\ \bibnamefont {Thorne}},
  \bibinfo {author} {\bibfnamefont {R.~W.}\ \bibnamefont {Drever}}, \bibinfo
  {author} {\bibfnamefont {V.~D.}\ \bibnamefont {Sandberg}}, \ and\ \bibinfo
  {author} {\bibfnamefont {M.}~\bibnamefont {Zimmermann}},\ }\href@noop {}
  {\bibfield  {journal} {\bibinfo  {journal} {Rev. Mod. Phys.}\ }\textbf
  {\bibinfo {volume} {52}},\ \bibinfo {pages} {341} (\bibinfo {year}
  {1980})}\BibitemShut {NoStop}%
\bibitem [{\citenamefont {Bocko}\ and\ \citenamefont {Onofrio}(1996)}]{Boc}%
  \BibitemOpen
  \bibfield  {author} {\bibinfo {author} {\bibfnamefont {M.~F.}\ \bibnamefont
  {Bocko}}\ and\ \bibinfo {author} {\bibfnamefont {R.}~\bibnamefont
  {Onofrio}},\ }\href@noop {} {\bibfield  {journal} {\bibinfo  {journal} {Rev.
  Mod. Phys.}\ }\textbf {\bibinfo {volume} {68}},\ \bibinfo {pages} {755}
  (\bibinfo {year} {1996})}\BibitemShut {NoStop}%
\bibitem [{\citenamefont {Caves}\ \emph {et~al.}(2012)\citenamefont {Caves},
  \citenamefont {Combes}, \citenamefont {Jiang},\ and\ \citenamefont
  {Pandey}}]{Cave12}%
  \BibitemOpen
  \bibfield  {author} {\bibinfo {author} {\bibfnamefont {C.~M.}\ \bibnamefont
  {Caves}}, \bibinfo {author} {\bibfnamefont {J.}~\bibnamefont {Combes}},
  \bibinfo {author} {\bibfnamefont {Z.}~\bibnamefont {Jiang}}, \ and\ \bibinfo
  {author} {\bibfnamefont {S.}~\bibnamefont {Pandey}},\ }\href {\doibase
  10.1103/PhysRevA.86.063802} {\bibfield  {journal} {\bibinfo  {journal} {Phys.
  Rev. A}\ }\textbf {\bibinfo {volume} {86}},\ \bibinfo {pages} {063802}
  (\bibinfo {year} {2012})}\BibitemShut {NoStop}%
\bibitem [{\citenamefont {Bassi}\ and\ \citenamefont {Ghirardi}(2003)}]{Bassi}%
  \BibitemOpen
  \bibfield  {author} {\bibinfo {author} {\bibfnamefont {A.}~\bibnamefont
  {Bassi}}\ and\ \bibinfo {author} {\bibfnamefont {G.}~\bibnamefont
  {Ghirardi}},\ }\href {\doibase
  http://dx.doi.org/10.1016/S0370-1573(03)00103-0} {\bibfield  {journal}
  {\bibinfo  {journal} {Phys. Rep.}\ }\textbf {\bibinfo {volume} {379}},\
  \bibinfo {pages} {257 } (\bibinfo {year} {2003})}\BibitemShut {NoStop}%
\bibitem [{\citenamefont {Kofler}\ and\ \citenamefont
  {Brukner}(2007)}]{Kofler}%
  \BibitemOpen
  \bibfield  {author} {\bibinfo {author} {\bibfnamefont {J.}~\bibnamefont
  {Kofler}}\ and\ \bibinfo {author} {\bibfnamefont {C.}~\bibnamefont
  {Brukner}},\ }\href {\doibase 10.1103/PhysRevLett.99.180403} {\bibfield
  {journal} {\bibinfo  {journal} {Phys. Rev. Lett.}\ }\textbf {\bibinfo
  {volume} {99}},\ \bibinfo {pages} {180403} (\bibinfo {year}
  {2007})}\BibitemShut {NoStop}%
\bibitem [{\citenamefont {Zhuang}\ \emph {et~al.}(2019)\citenamefont {Zhuang},
  \citenamefont {Schuster}, \citenamefont {Yoshida},\ and\ \citenamefont
  {Yao}}]{Yao}%
  \BibitemOpen
  \bibfield  {author} {\bibinfo {author} {\bibfnamefont {Q.}~\bibnamefont
  {Zhuang}}, \bibinfo {author} {\bibfnamefont {T.}~\bibnamefont {Schuster}},
  \bibinfo {author} {\bibfnamefont {B.}~\bibnamefont {Yoshida}}, \ and\
  \bibinfo {author} {\bibfnamefont {N.~Y.}\ \bibnamefont {Yao}},\ }\href@noop
  {} {\bibfield  {journal} {\bibinfo  {journal} {arXiv preprint
  arXiv:1902.04076}\ } (\bibinfo {year} {2019})}\BibitemShut {NoStop}%
\bibitem [{\citenamefont {Rossi}\ \emph {et~al.}(2017)\citenamefont {Rossi},
  \citenamefont {Foti}, \citenamefont {Cuccoli}, \citenamefont {Trapani},
  \citenamefont {Verrucchi},\ and\ \citenamefont {Paris}}]{Rossi}%
  \BibitemOpen
  \bibfield  {author} {\bibinfo {author} {\bibfnamefont {M.~A.}\ \bibnamefont
  {Rossi}}, \bibinfo {author} {\bibfnamefont {C.}~\bibnamefont {Foti}},
  \bibinfo {author} {\bibfnamefont {A.}~\bibnamefont {Cuccoli}}, \bibinfo
  {author} {\bibfnamefont {J.}~\bibnamefont {Trapani}}, \bibinfo {author}
  {\bibfnamefont {P.}~\bibnamefont {Verrucchi}}, \ and\ \bibinfo {author}
  {\bibfnamefont {M.~G.}\ \bibnamefont {Paris}},\ }\href@noop {} {\bibfield
  {journal} {\bibinfo  {journal} {Phys. Rev. A}\ }\textbf {\bibinfo {volume}
  {96}},\ \bibinfo {pages} {032116} (\bibinfo {year} {2017})}\BibitemShut
  {NoStop}%
\bibitem [{\citenamefont {Mollow}\ and\ \citenamefont
  {Glauber}(1967{\natexlab{a}})}]{Mol1}%
  \BibitemOpen
  \bibfield  {author} {\bibinfo {author} {\bibfnamefont {B.}~\bibnamefont
  {Mollow}}\ and\ \bibinfo {author} {\bibfnamefont {R.}~\bibnamefont
  {Glauber}},\ }\href@noop {} {\bibfield  {journal} {\bibinfo  {journal} {Phys.
  Rev.}\ }\textbf {\bibinfo {volume} {160}},\ \bibinfo {pages} {1076} (\bibinfo
  {year} {1967}{\natexlab{a}})}\BibitemShut {NoStop}%
\bibitem [{\citenamefont {Mollow}\ and\ \citenamefont
  {Glauber}(1967{\natexlab{b}})}]{Mol2}%
  \BibitemOpen
  \bibfield  {author} {\bibinfo {author} {\bibfnamefont {B.~R.}\ \bibnamefont
  {Mollow}}\ and\ \bibinfo {author} {\bibfnamefont {R.~J.}\ \bibnamefont
  {Glauber}},\ }\href {\doibase 10.1103/PhysRev.160.1097} {\bibfield  {journal}
  {\bibinfo  {journal} {Phys. Rev.}\ }\textbf {\bibinfo {volume} {160}},\
  \bibinfo {pages} {1097} (\bibinfo {year} {1967}{\natexlab{b}})}\BibitemShut
  {NoStop}%
\bibitem [{\citenamefont {Collett}\ and\ \citenamefont {Walls}(1988)}]{Col}%
  \BibitemOpen
  \bibfield  {author} {\bibinfo {author} {\bibfnamefont {M.}~\bibnamefont
  {Collett}}\ and\ \bibinfo {author} {\bibfnamefont {D.}~\bibnamefont
  {Walls}},\ }\href@noop {} {\bibfield  {journal} {\bibinfo  {journal} {Phys.
  Rev. Lett.}\ }\textbf {\bibinfo {volume} {61}},\ \bibinfo {pages} {2442}
  (\bibinfo {year} {1988})}\BibitemShut {NoStop}%
\bibitem [{\citenamefont {Lee}\ and\ \citenamefont {Jeong}(2011)}]{Lee}%
  \BibitemOpen
  \bibfield  {author} {\bibinfo {author} {\bibfnamefont {C.-W.}\ \bibnamefont
  {Lee}}\ and\ \bibinfo {author} {\bibfnamefont {H.}~\bibnamefont {Jeong}},\
  }\href {\doibase 10.1103/PhysRevLett.106.220401} {\bibfield  {journal}
  {\bibinfo  {journal} {Phys. Rev. Lett.}\ }\textbf {\bibinfo {volume} {106}},\
  \bibinfo {pages} {220401} (\bibinfo {year} {2011})}\BibitemShut {NoStop}%
\end{thebibliography}
\end{document}